\documentclass[aps,prapplied,reprint,superscriptaddress,nofootinbib]{revtex4-2}
\usepackage[utf8]{inputenc}
\usepackage{amsmath,amssymb,amsfonts}
\usepackage{physics}
\usepackage{braket}
\usepackage{graphicx}
\usepackage{booktabs}
\usepackage{algorithm}
\usepackage{algpseudocode}
\usepackage{hyperref}
\hypersetup{
	colorlinks=true,
	linkcolor=blue,
	citecolor=blue,
	urlcolor=blue,
}
\usepackage{xcolor}
\usepackage{array}
\usepackage{listings}
\usepackage{adjustbox}
\usepackage{multirow}
\usepackage{standalone}
\usepackage{float}   
\usepackage{quantikz}
\usepackage{tabularx}
\usepackage{siunitx}

\newcommand{\LAU}{Department of Electrical and Computer Engineering, Lebanese American University (LAU), P.O. Box 36, Byblos, Lebanon}
\newcommand{\BMW}{BMW Group, Munich, Germany}
\newcommand{\LMU}{Ludwig Maximilian University, Munich, Germany}

\newif\ifdraft
\drafttrue
\ifdraft
\newcommand{\alnote}[1]{ {\textcolor{purple} { ***Andre: #1 }}}
\newcommand{\fknote}[1]{ {\textcolor{blue} { ***Florian: #1 }}}
\newcommand{\crnote}[1]{ {\textcolor{orange} { ***Carlos: #1 }}}
\newcommand{\kanote}[1]{ {\textcolor{red} { ***Kevin: #1 }}}
\else
\newcommand{\alnote}[1]{}
\newcommand{\fknote}[1]{}
\newcommand{\crnote}[1]{}
\newcommand{\kanote}[1]{}
\fi

\begin{document}


\title{Quantum State Preparation via Neural Network Encoding\\ in Quantum Machine Learning}

\author{Kevin W. Aoun}
\email{kevin.aoun02@lau.edu}
\affiliation{\LAU}

\author{Florian J. Kiwit}
\email{florian.kiwit@ifi.lmu.de}
\affiliation{\BMW}
\affiliation{\LMU}

\author{Carlos A. Riofr\'io}
\affiliation{\BMW}

\author{Samer Saab Jr.}
\affiliation{\LAU}

\author{Charbel Al Bateh}
\affiliation{\LAU}

\author{Joe Tekli}
\affiliation{\LAU}

\author{Andre Luckow}
\affiliation{\BMW}
\affiliation{\LMU}

\begin{abstract}
A central challenge in quantum machine learning is the state preparation bottleneck that describes the prohibitive computational cost of loading high-dimensional classical data into a quantum state. Although amplitude encoding can represent $2^n$-dimensional data using only $n$ qubits in principle,
preparing arbitrary states remains computationally expensive, typically requiring variational optimization of a parameterized quantum circuit for each individual data instance. In this work, we propose a method that avoids iterative optimization by training a classical neural network to map input data directly to the continuous parameters of a fixed quantum circuit. We demonstrate the generation of quantum image states with high fidelity on data not seen during training. Since all optimization is performed once during training, the resulting model encodes new inputs in a single inference step, providing a scalable pathway for data loading in near-term quantum algorithms. We validate our method on the MNIST and Fashion-MNIST datasets, achieving fidelities up to 0.992 on unseen images and reducing the per-data-instance runtime by more than 5000-fold.
\end{abstract}

\maketitle

\section{Introduction}
\label{sec:intro}

Quantum computing promises computational advantages over classical approaches, with quantum algorithms demonstrating theoretical speedups for problems ranging from integer factorization~\cite{shor1994algorithms} to unstructured search~\cite{grover1996fast} and linear system solvers~\cite{harrow2009quantum}. Beyond these landmark results, quantum machine learning (QML) has emerged as a promising direction, with theoretical evidence suggesting advantages for certain learning tasks in specific regimes~\cite{biamonte2017quantum,rebentrost2014quantum,Liu_2021}. However, a critical prerequisite for all such algorithms is that the input data is already prepared as a quantum state. When the cost of loading classical data into quantum states is included in the overall algorithmic complexity, any speedup offered by the subsequent quantum algorithm may already be negated at the data loading stage~\cite{aaronson2015read}.

This state preparation bottleneck is particularly limiting for QML on near-term devices. Encoding schemes such as amplitude encoding~\cite{schuld2022machine,latorre2005imagecompressionentanglement}, where the data is encoded into the amplitudes of a superposition of basis states, in principle allow encoding an exponential number of data components in a linear number of qubits. The caveat is that preparing arbitrary data vectors using this encoding becomes exponentially costly~\cite{schuld2022machine}.

However, actual data of interest is often not arbitrary but has some internal structure, e.g., how images differ from random pixel values. In fact, several recent results (both numerical and theoretical) show that typical images can be well approximated by quantum states~\cite{Dilip2022, iaconis2023tensornetworkbasedefficient, jobst2024efficient, shen2024classification,maxwell_2024}. These states can be prepared using tensor-network-inspired circuits whose depth scales only linearly in the number of qubits. While we focus on image data in this paper, there are other types of data which result in lowly entangled states after amplitude encoding, for which we expect the methods of this paper to work similarly well~\cite{Lubasch2018, Lubasch2020, Gourianov2022, Hoelscher2025, Ritter2024, jobst2024efficient}. Building on these insights, Ref.~\cite{kiwit2025typical} developed a practical optimization pipeline to generate low-depth circuits for standard machine learning datasets. While this approach demonstrates that realistic datasets admit efficient circuit representations, it still requires solving a classical optimization problem separately for each individual input. This per-instance optimization becomes a bottleneck if one ultimately seeks a state-preparation routine that can be deployed repeatedly within a larger QML pipeline.  

A related line of work replaces per-instance optimization with learned circuit generation. Most notably, Daimon and Matsushita~\cite{daimon2024quantum} formulated amplitude encoding as a sequence-generation problem. They trained a transformer decoder that takes a tokenized target amplitude vector as input and autoregressively generates a tokenized quantum gate sequence. They report strong generalization to unseen inputs, high test-state fidelity, and in some cases circuits that are even shallower than those used during training, suggesting that transformer models can learn nontrivial structural regularities in state-preparation problems. In contrast, our approach fixes the circuit structure upfront and trains a neural network to predict its rotation angles directly from raw image data, requiring no gate-sequence generation and no pre-computed amplitude vectors.

In this work, we present such a form of learned state preparation: rather than generating a flexible gate sequence or optimizing a circuit individually for every sample, we train a classical neural network to map an input image directly to the continuous parameters of a fixed parameterized quantum circuit (PQC). The PQC then encodes the image using the flexible representation of quantum images (FRQI)~\cite{le2011flexible,le2011flexible2}. By offloading the optimization cost into an offline training phase, the resulting framework produces the circuit representation for new inputs in a single inference step, eliminating the need for iterative optimization at inference time. Our key contributions are (i) a quantum-classical architecture that maps input images to gate parameters encoding FRQI states, (ii) a systematic analysis of ansatz designs and circuit depths characterizing the trade-off between circuit complexity and state fidelity, and (iii) a demonstration of the speedup of our method over previous state-preparation approaches.

The paper is organized as follows. Section~\ref{sec:representation} introduces the basic concepts of image representation as quantum circuits. Section~\ref{sec:methods} presents the proposed framework and experimental design. Section~\ref{sec:results} provides a comprehensive evaluation of reconstruction quality and runtime, including ablation studies on parameter bounding and circuit architecture. Finally, Section~\ref{sec:conclusion} summarizes our findings and discusses future directions.

\section{Quantum Image Representations}
\label{sec:representation}

A natural strategy for embedding classical data into a quantum system is \emph{amplitude encoding}, in which a data vector $\mathbf{x}$ is mapped to the state $\ket{\psi(\mathbf{x})} = \tfrac{1}{\|\mathbf{x}\|} \sum_{j=0}^{2^A - 1} x_j \ket{j}$~\cite{schuld2022machine,latorre2005imagecompressionentanglement}. Because only $A$ qubits are needed to represent $2^A$ pixel values, this scheme achieves an exponential compression. A limitation, however, is that the classical data vector must be normalized to form a valid quantum state, which removes information about its global scale.

To preserve per-pixel intensity while retaining exponential compression, the \emph{Flexible Representation of Quantum Images}~(FRQI)~\cite{le2011flexible,le2011flexible2} has been introduced. For an image with $2^A$ pixels, the quantum state is given by
\begin{equation}
    \label{eq:target_state}
    \ket{\psi(\mathbf{x})} = \frac{1}{\sqrt{2^A}} \sum_{j=0}^{2^A-1} \ket{c(\mathbf{x}_j)} \otimes \ket{j}
    ,
\end{equation}
where $\ket{c(x_j)} = \cos\!\bigl({\textstyle\frac{\pi}{2}}\, x_j\bigr)\, \ket{0} + \sin\!\bigl({\textstyle\frac{\pi}{2}}\, x_j\bigr)\, \ket{1}$. Here, the state $\ket{j}$ of the $A$ so-called \emph{address qubits} encodes the pixel position, while the single \emph{color qubit} state $\ket{c(x_j)}$ encodes the corresponding grayscale intensity $x_j \in [0,1]$. Thus, an image with $2^A$ pixels can be represented using $n = A+1$ qubits.

\begin{figure}[!tbp]
    \centering
    \includegraphics[width=0.4\textwidth]{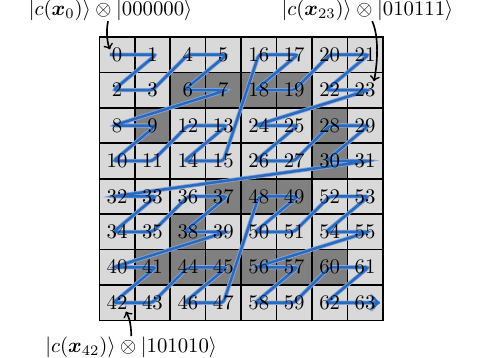}
    \caption{Hierarchical (Z-order) pixel indexing. Pixels are traversed by recursively partitioning the image into quadrants. This mapping reduces the entanglement entropy of the target state.}
    \label{fig:z_order}
\end{figure}

\paragraph{Hierarchical (Z-order) indexing.}
The pixel ordering used to assign indices $j$ can affect the entanglement structure of the resulting state~\cite{jobst2024efficient}. A conventional row-based scan destroys two-dimensional spatial correlations and typically leads to higher entanglement entropy, making the state harder to approximate with low-depth or linear ans\"atze such as the MPS circuit used in this work. We therefore adopt \emph{hierarchical indexing} based on the Morton (Z-order) curve~\cite{morton1966computer,latorre2005imagecompressionentanglement,le2011flexible,le2011flexible2,jobst2024efficient}: the two most-significant bits of $j$ identify the image quadrant, the next two identify the sub-quadrant, and so on recursively, as shown in Fig.~\ref{fig:z_order}. This ordering has been shown to achieve lower entanglement entropies~\cite{kiwit2025typical,jobst2024efficient}.

\paragraph{Decoding and reconstruction sensitivity.}
Classical pixel values can be recovered from the color-qubit amplitudes conditioned on each address state. For a given address state $\ket{j}$, let $a_{0,j}$ and $a_{1,j}$ denote the conditional $\ket{0}$- and $\ket{1}$-amplitudes of the corresponding color-qubit state, respectively.
The grayscale intensity encoded at pixel position $j$ can then be reconstructed as
\begin{equation}
    \label{eq:decoding}
    \tilde{x}_j
    =
    \frac{2}{\pi}\arccos(a_{0,j})
    =
    \frac{2}{\pi}\arcsin(a_{1,j}).
\end{equation}

\section{Methods}
\label{sec:methods}

\begin{figure*}[!htbp]
    \centering
    \includegraphics[width=1.0\textwidth]{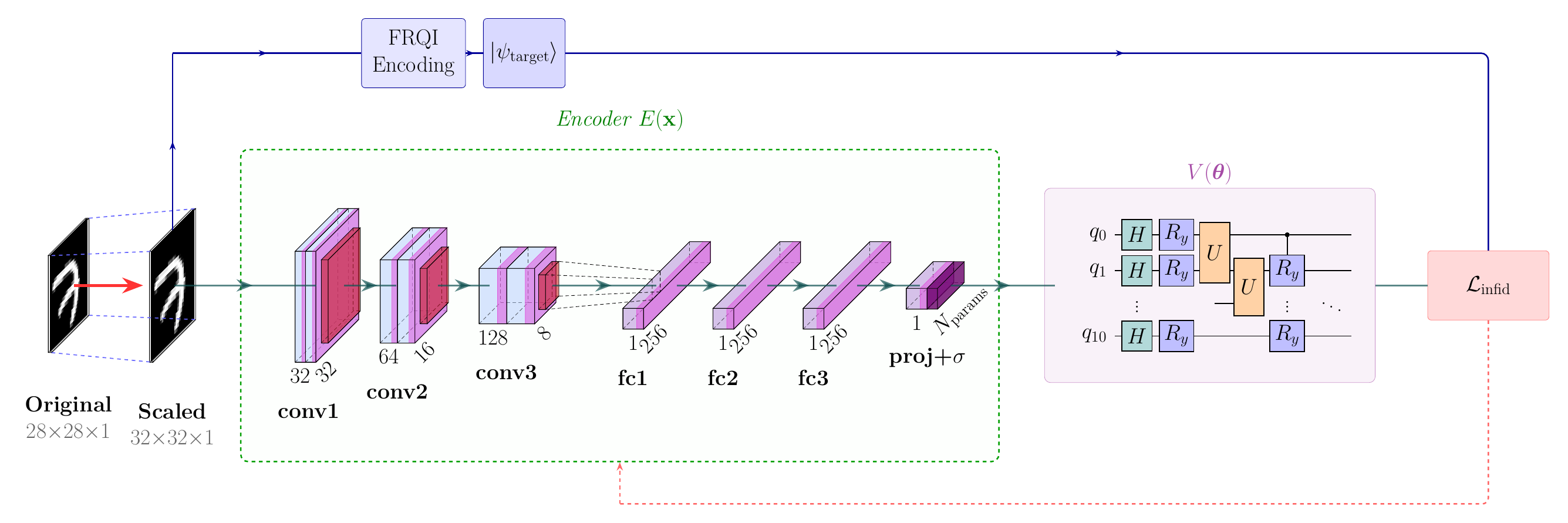}
    \caption{\textbf{Direct state preparation framework.} Input images $\mathbf{x}$ are encoded via a CNN $E(\mathbf{x})$ into PQC parameters $\boldsymbol{\theta} \in \mathbb{R}^{N_{\text{params}}}$. The circuit generates a quantum state $\ket{\psi(\mathbf{\theta})}$ trained to match the ground-truth FRQI state, $\ket{\psi_{target}}$, via an infidelity loss.}
    \label{fig:architecture}
\end{figure*}

The core of our approach is a convolutional neural network (CNN) encoder $E(\mathbf{x}) = \boldsymbol{\theta}$ that maps an input image $\mathbf{x}$ to a vector of parameterized quantum circuit (PQC) parameters $\boldsymbol{\theta} \in \mathbb{R}^{N_{\text{params}}}$, as depicted in Fig.~\ref{fig:architecture}. The encoder consists of three successive Conv--Swish--MaxPool blocks with 32, 64, and 128 filters ($3\times3$ kernels), transforming a $32\times32$ grayscale image to a $4\times4\times128$ feature map before flattening to a 2048-dimensional vector. A three-layer MLP with 256 hidden units and Swish activations follows, giving a backbone of approximately 749{,}000 trainable weights independent of the ansatz. A final linear projection maps to $\boldsymbol{\theta}$, initialized with small variance scaling to prevent large initial rotation angles. In the bounded variant, a sigmoid nonlinearity constrains each output to $[0, 2\pi)$.

It is important to distinguish between two types of parameters in the framework: the weights of the encoder $E(x)$, which are fixed after training, and the circuit parameters $\boldsymbol{\theta}$, which are the output of the neural network and vary with each input image. 

\subsection{Quantum circuit architectures}
\begin{figure*}[!htbp]
\centering
\includegraphics[width=\textwidth]{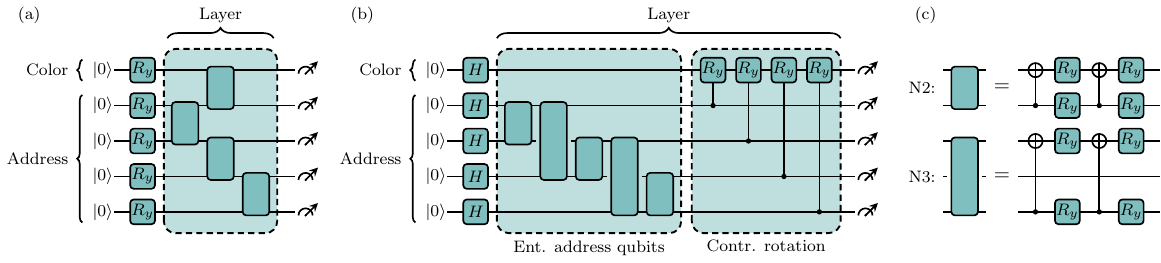}

\caption{\textbf{Architectural designs for the quantum generator and constituent gate decompositions.} 
(a) center-sequential circuit: an integrated layer design where the color qubit and address register are entangled through a sequential arrangement of multi-qubit blocks. 
(b) FRQI-specific circuit: a structured layer implementation for a $4 \times 4$-pixel image. The process begins with Hadamard ($H$) gates preparing an initial uniform superposition across all qubits. The layer is partitioned into two functional blocks: an entangling address-qubits block, which utilizes alternating N2 and N3 blocks to create correlations within the address register, and controlled rotations, where the color qubit undergoes $R_y$ rotations conditioned on the state of each address qubit to map spatial coordinates to pixel intensities. 
(c) The right panel illustrates the gate decompositions for the N2 and N3 primitive blocks, showing how these operations are implemented using standard CNOT gates and parameterized $R_y$ rotations.}
\label{fig:ansatz_comparison}
\end{figure*}

We define two architectures for the parameterized quantum circuit as depicted in Fig.~\ref{fig:ansatz_comparison}. Given the predicted parameters, the PQC $V(\boldsymbol{\theta})$ prepares the quantum state $\ket{\psi(\boldsymbol{\theta})} = V(\boldsymbol{\theta})\ket{0}^{\otimes n}$.

\paragraph{Center-sequential circuit.}
Sequential quantum circuits are widely used as circuit architectures as they are closely related to matrix product states (MPS)~\cite{fannes1992finitely,cirac2021matrix}. One layer applies gates on neighboring qubits in a staircase pattern from the first to the last qubit register. To reduce the circuit depth, one can change from the sequential circuit layer to a center-sequential layer~\cite{kiwit2025typical}, where the sequential arms fan out from the central qubit register as depicted in Fig.~\ref{fig:ansatz_comparison}(a). This yields a parameter count of $ N_\text{params}=4(n-1)D + n$. For $n = 11$ qubits at depth $D = 6$, this translates to $N_\text{params} = 251$. 

\paragraph{FRQI-native circuit.}
The second ansatz, shown in Fig.~\ref{fig:ansatz_comparison}(b), incorporates an inductive bias for the structure of the FRQI target state, as shown in Eq.~\eqref{eq:target_state}~\cite{jaeger2026scalingquantummachinelearning}. After initializing the qubits in a uniform superposition via Hadamard gates, it applies layers that consist of two parts. First, the address qubits are entangled via nearest-neighbor (N2) and next-nearest-neighbor (N3) gates. Second, the $R_y$ rotations controlled by the address qubits are applied. For $D$ layers, this yields a parameter count of $N_{\text{params}} = (9n - 21)D$. For $n = 11$ qubits at depth $D = 6$, this gives $N_{\text{params}} = 468$.

\subsection{Datasets}
\label{sec:datasets}

We evaluate on two standard benchmarks, the MNIST and Fashion-MNIST datasets. Since FRQI encoding requires the pixel count to be a power of two ($N = 2^A$), we resize all images from $28 \times 28$ to $32 \times 32$ pixels ($N = 1024$, $A = 10$) using bilinear interpolation, yielding $n = A + 1 = 11$ qubits in total. The MNIST dataset~\cite{lecun1998gradient} contains \num{70000} handwritten digit images ($28\times28$ pixels) across 10 classes, split into \num{60000} training and \num{10000} test samples. Fashion-MNIST~\cite{xiao2017fashion} shares the same format and split but depicts clothing items (T-shirt, trouser, pullover, dress, coat, sandal, shirt, sneaker, bag, ankle boot), providing a more structurally complex test of generalization. Both datasets undergo identical preprocessing: bilinear resizing to $32\times32$, intensity normalization to $[0,1]$, and FRQI encoding with hierarchical (Z-order) indexing.

\subsection{Evaluation metrics}
\label{sec:metrics}

We assess state preparation quality and image reconstruction through three complementary metrics.

\paragraph{State fidelity.}
The overlap between the prepared state and the FRQI ground truth measures how faithfully the circuit reproduces the target quantum state:
\begin{equation}
    \label{eq:fidelity}
    \mathcal{F} = \bigl|\braket{\psi_{\text{target}}|\psi(\boldsymbol{\theta})}\bigr|^2 .
\end{equation}
Training minimizes the \emph{infidelity} loss $\mathcal{L} = 1 - \mathcal{F}$.

\paragraph{Peak signal-to-noise ratio (PSNR).}
After decoding the prepared state back to pixel intensities via Eq.~\eqref{eq:decoding}, PSNR quantifies the pixel-level reconstruction accuracy on a logarithmic scale:
\begin{equation}
    \label{eq:psnr}
    \text{PSNR} = 10\,\log_{10}\!\left(\frac{MAX^2}{\text{MSE}}\right), \ 
    \text{MSE} = \frac{1}{N}\sum_{j=0}^{N-1}\bigl(x_j - \tilde{x}_j\bigr)^2 ,
\end{equation}
where $x_j$ and $\tilde{x}_j$ are the original and reconstructed pixel values, respectively.
Since pixel intensities are normalized to $[0,1]$, the peak signal value is $\mathrm{MAX} = 1$.

\paragraph{Structural similarity index (SSIM).}
SSIM~\cite{wang2004image} evaluates perceptual reconstruction quality by comparing local luminance, contrast, and structural patterns between the original image $\mathbf{x}$ and its reconstruction $\tilde{\mathbf{x}}$:
\begin{equation}
    \label{eq:ssim}
    \text{SSIM}(\mathbf{x}, \tilde{\mathbf{x}}) = \frac{(2\mu_x \mu_{\tilde{x}} + c_1)(2\sigma_{x\tilde{x}} + c_2)}{(\mu_x^2 + \mu_{\tilde{x}}^2 + c_1)(\sigma_x^2 + \sigma_{\tilde{x}}^2 + c_2)} ,
\end{equation}
where $\mu$, $\sigma^2$, and $\sigma_{x\tilde{x}}$ denote local means, variances, and cross-covariance, and $c_1 = (k_1 L)^2$, $c_2 = (k_2 L)^2$ are stabilization constants with $k_1 = 0.01$, $k_2 = 0.03$, and $L$ the dynamic range of the pixel values~\cite{wang2004image}. Unlike PSNR, SSIM captures perceptual degradation that pixel-wise metrics may miss.

\paragraph{Relationship between metrics.}
Fidelity operates in Hilbert space and captures global agreement of the quantum state amplitudes, whereas PSNR and SSIM operate on the decoded image and are sensitive to the nonlinear arccosine inversion, as shown in Eq.~\eqref{eq:decoding}. High fidelity does not guarantee high PSNR or SSIM, and vice versa; we therefore report all three to provide a complete picture of preparation quality.

\subsection{Experimental protocol}
\label{sec:protocol}

All experiments follow a controlled protocol designed to isolate the effect of each architectural choice. We vary one factor at a time (ansatz type, circuit depth $D$, and dataset) while holding the remaining hyperparameters fixed. For each configuration we record: (i)~infidelity curves, (ii)~PSNR and SSIM on decoded reconstructions, and (iii)~total trainable parameter count $K$. 
For evaluation, we randomly sample 2048 images from each test set to reduce the computational cost of statevector simulation during metric computation; all training uses the full training split.
The models, implemented with PennyLane~\cite{bergholm2018pennylane} (JAX backend) for circuit simulation and JAX/Flax for the classical encoder, are trained using the Adam optimizer with a learning rate of $10^{-3}$ on an NVIDIA A100 GPU. Each run is repeated with three random seeds and we report mean and standard deviation.

\section{Results}
\label{sec:results}

\begin{figure*}[!htbp]
    \centering
    \includegraphics[width=\textwidth]{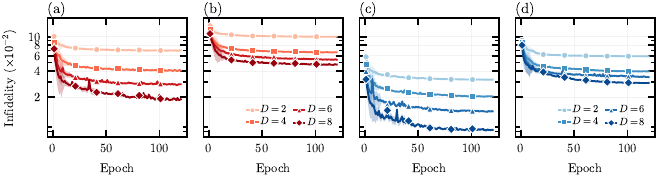}
    \caption{Validation infidelity vs.\ training epoch for circuit depths $D \in \{2, 4, 6, 8\}$. Red curves correspond to MNIST and blue curves to Fashion-MNIST; within each color family, darker shades indicate greater depth. (a)~MNIST, FRQI ansatz. (b)~MNIST, center-sequential ansatz. (c)~Fashion-MNIST, FRQI ansatz. (d)~Fashion-MNIST, center-sequential ansatz. Both architectures show consistent improvement with depth. F-MNIST converges to lower infidelity than MNIST at equivalent depth, though the FRQI ansatz at $D \in \{6, 8\}$ ($N_{\text{params}} = 468$ and $624$) exhibits some training instability on both datasets.}
    \label{fig:depth_sweep}
\end{figure*}

In the following, we evaluate the proposed framework on the MNIST and Fashion-MNIST datasets using both circuit architectures and circuit depths
$D \in \{2,4,6,8\}$ in terms of quality and speed.

\paragraph{State-preparation quality.}
Figure~\ref{fig:depth_sweep} shows the validation infidelity as a function of training epoch across both datasets and circuit architectures. All configurations converge within the first ${\sim}20$ epochs, with the strongest performance gain occurring between $D{=}2$ and $D{=}4$. Beyond $D{=}6$, improvements plateau: the FRQI ansatz improves from a fidelity of $0.972$ to $0.981$ on MNIST, while the center-sequential ansatz only increases marginally from $0.946$ to $0.952$. Both circuit architectures exhibit training instability at higher depths, though this effect is more pronounced for the FRQI ansatz at $D \in \{6,8\}$, consistent with the larger parameter count of this architecture ($468$--$624$ versus $251$--$331$ for center-sequential).

\begin{figure}[t]
    \centering
    \includegraphics[width=\columnwidth]{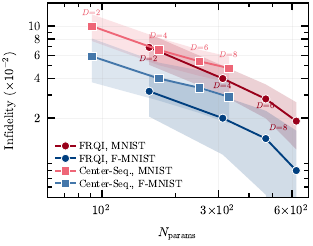}
    \caption{Scaling of the infidelity with an increase in circuit parameters $N_{\text{params}}$ for the FRQI and center-sequential ans\"atze on MNIST and Fashion-MNIST. Each marker corresponds to a circuit depth $D \in \{2, 4, 6, 8\}$. The shaded areas show the 25th-75th percentiles, scaled down for visual purposes. At matched parameter counts, FRQI consistently achieves higher fidelity, confirming that the gain is architectural rather than a consequence of increased model capacity. Red denotes MNIST and blue denotes Fashion-MNIST; darker shades correspond to the FRQI ansatz and lighter shades to the center-sequential ansatz.}
    \label{fig:fidelity_vs_params}
\end{figure}

Table~\ref{tab:depth_results} presents the fidelity, PSNR, and SSIM of all configurations after $100$ training epochs, evaluated on $2048$ held-out samples. The FRQI ansatz outperforms the center-sequential ansatz at every depth on both datasets. To separate the effect of ansatz architecture from parameter count, Fig.~\ref{fig:fidelity_vs_params} plots fidelity against $N_{\text{params}}$ for both ans\"atze. At comparable parameter budgets, FRQI consistently achieves higher fidelity: for example, FRQI at $D{=}4$ with $312$ parameters reaches a fidelity of $0.960$ on MNIST, surpassing center-sequential at $D{=}8$ with $331$ parameters and fidelity of $0.952$. This confirms that the FRQI advantage is architectural: its inductive bias toward the FRQI target-state structure yields higher fidelity per parameter, rather than the improvement arising from increased parameter count alone. Additionally, Fashion-MNIST achieves higher fidelity than MNIST at every configuration (e.g., $0.986$ vs.\ $0.972$ for FRQI at $D{=}6$), likely because fashion items contain smoother intensity distributions. However, higher fidelity does not necessarily imply better visual reconstruction: Fashion-MNIST yields lower PSNR and SSIM because its images contain high-frequency textures such as textile patterns and structured details, which the model smooths while preserving overall shape and class identity.

\begin{figure*}[!htbp]
    \centering
    \includegraphics[width=\textwidth]
    {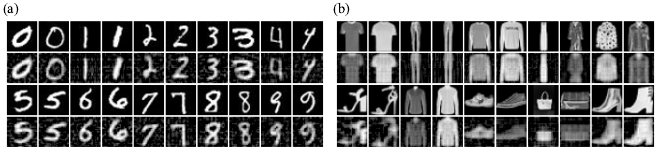}
    \caption{Image reconstructions from FRQI encoded samples of (a) MNIST and (b) Fashion-MNIST datasets with depth $D=8$. The first and third row show original test images and the second and fourth rows show the corresponding reconstructions. The model captures dominant shape and intensity patterns, but struggles to capture fine details.
    }
    \label{fig:reconstructions}
\end{figure*}

Figure~\ref{fig:reconstructions} displays image reconstructions from the best-performing configuration (FRQI, $D{=}8$) on both datasets. The encoder captures dominant shape and intensity patterns, but finer details degrade for visually complex categories such as digit loops and textured garments.

\begin{table*}[!htbp]
    \centering
    \caption{Reconstruction metrics for varying ansatz and circuit depth on MNIST and Fashion-MNIST ($n=11$ qubits, 2048 held-out samples).}
    \label{tab:depth_results}
    \setlength{\tabcolsep}{0pt}
    \begin{tabular*}{\textwidth}{@{\extracolsep{\fill}} lcc ccc ccc}
        \toprule
        & & & \multicolumn{3}{c}{\textbf{MNIST}} & \multicolumn{3}{c}{\textbf{Fashion-MNIST}} \\
        \cmidrule(lr){4-6} \cmidrule(lr){7-9}
        \textbf{Ansatz} & \textbf{Depth} & $N_{\text{params}}$
          & \textbf{Fid.} & \textbf{PSNR} & \textbf{SSIM}
          & \textbf{Fid.} & \textbf{PSNR} & \textbf{SSIM} \\
        \midrule
        \multirow{4}{*}{Center-Seq.}
        & 2 &  91
          & $0.900 \pm 0.034$ & $13.23 \pm 1.35$ & $0.345 \pm 0.059$
          & $0.941 \pm 0.032$ & $15.95 \pm 2.24$ & $0.437 \pm 0.101$ \\
        & 4 & 171
          & $0.934 \pm 0.023$ & $14.33 \pm 1.12$ & $0.414 \pm 0.078$
          & $0.960 \pm 0.019$ & $17.00 \pm 1.73$ & $0.477 \pm 0.090$ \\
        & 6 & 251
          & $0.946 \pm 0.018$ & $14.85 \pm 0.99$ & $0.447 \pm 0.088$
          & $0.966 \pm 0.018$ & $16.99 \pm 1.43$ & $0.476 \pm 0.101$ \\
        & 8 & 331
          & $\mathbf{0.952 \pm 0.016}$ & $\mathbf{15.04 \pm 0.80}$ & $\mathbf{0.465 \pm 0.098}$
          & $\mathbf{0.971 \pm 0.016}$ & $\mathbf{17.70 \pm 1.58}$ & $\mathbf{0.513 \pm 0.097}$ \\
        \midrule
        \multirow{4}{*}{FRQI}
        & 2 & 156
          & $0.931 \pm 0.027$ & $14.09 \pm 1.22$ & $0.396 \pm 0.080$
          & $0.968 \pm 0.017$ & $17.55 \pm 1.70$ & $0.500 \pm 0.096$ \\
        & 4 & 312
          & $0.960 \pm 0.015$ & $15.43 \pm 0.87$ & $0.482 \pm 0.097$
          & $0.980 \pm 0.014$ & $18.65 \pm 1.60$ & $0.561 \pm 0.107$ \\
        & 6 & 468
          & $0.972 \pm 0.012$ & $16.20 \pm 0.89$ & $0.517 \pm 0.103$
          & $0.986 \pm 0.013$ & $19.31 \pm 1.63$ & $0.590 \pm 0.111$ \\
        & 8 & 624
          & $\mathbf{0.981 \pm 0.011}$ & $\mathbf{16.89 \pm 0.86}$ & $\mathbf{0.543 \pm 0.105}$
          & $\mathbf{0.992 \pm 0.012}$ & $\mathbf{19.74 \pm 1.68}$ & $\mathbf{0.606 \pm 0.109}$ \\
        \bottomrule
    \end{tabular*}
\end{table*}

\paragraph{Runtime analysis.}\label{sec:runtime}
To quantify the practical advantage of our approach, we compare the wall-clock cost of state preparation against per-image classical optimization on the same hardware (CPU inference). As a baseline, we optimize an MPS circuit ($D{=}6$, $\chi_{\max}{=}256$) for each test image individually using the iterative sweeping procedure of Ref.~\cite{kiwit2025typical}, terminating early once a fidelity of $\mathcal{F} \geq 0.950$ is reached. Our method uses the MPS-inspired center-sequential ansatz targeting the same fidelity threshold; although the two circuit architectures differ in structure, matching $\mathcal{F}$ ensures a fair comparison at equivalent preparation quality. Averaged over 10 samples, the baseline requires $3.27$\,s per image, with substantial variance across samples driven by image-dependent convergence behavior: simpler images with low-rank structure converge in a handful of sweeps, while images with richer high-frequency content require many more before reaching the fidelity threshold.

In our method, the parameter prediction reduces to a single forward pass through the CNN encoder and no optimization is required at inference time. For the center-sequential ansatz with depth $D{=}6$, the median prediction latency is $0.497$\,ms per image, a $\sim6580$-fold speedup over per-image optimization.

\begin{figure}[!tbp]
    \centering
    \includegraphics[width=\columnwidth]{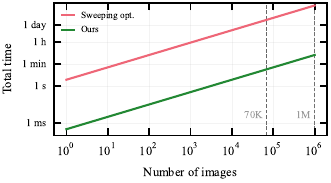}
    \caption{Total wall-clock time as a function of dataset size for per-image sweeping optimization ($3.27$\,s/image) versus amortized encoder inference ($\sim0.5$\,ms/image). Dashed lines mark the MNIST ($70$K) and industrial ($1$M) scales.}
    \label{fig:runtime}
\end{figure}

Figure~\ref{fig:runtime} translates these per-image costs to dataset scale. Preparing all 70\,000 MNIST images via sweeping would require ${\sim}2.65$\,days; our encoder completes the same task in ${\sim}35$\,seconds. At industrial scale ($10^6$ images), the gap widens from ${\sim}38$\,days to ${\sim}8.3$\,minutes, a difference that determines whether quantum data loading is deployable in practice.

\paragraph{Limitations}
All experiments used low-resolution ($32\times32$) grayscale benchmarks; scaling to higher resolutions would require deeper circuits or hierarchical encodings whose trainability remains open. The PQCs consistently struggle to reproduce high-frequency spatial components, producing smoothed reconstructions that limit applicability to fine-grained tasks. All simulations used ideal state-vector backends; hardware noise and finite shot effects have not been evaluated.

\section{Conclusion and Future Work}
\label{sec:conclusion}

We introduced a state-preparation framework that replaces iterative, per-instance circuit optimization with a neural network that maps input images to the parameters of a quantum circuit that encodes its FRQI representation. This shifts the computational cost from online state preparation to an offline training phase and enables new, unseen inputs to be encoded in a single forward pass. Our results show that this approach can generate high-fidelity FRQI states with shallow circuits. Using the FRQI-native ansatz, the model achieves average fidelities of $0.981$ on MNIST and $0.992$ on Fashion-MNIST on the test images. The comparison between ansatz families further shows that the FRQI-native circuit achieves higher fidelity per parameter than the center-sequential architecture, indicating that the improvement is not merely due to increased parameter count but to an architectural inductive bias aligned with the target quantum-image structure. The main practical advantage of the proposed method lies in the runtime. Compared with per-image sweeping optimization, which requires seconds per image, the trained encoder predicts circuit parameters in approximately 0.5 ms per image, corresponding to a speedup of more than three orders of magnitude.

While the current framework is limited to grayscale images, future iterations will extend the encoding to color images by using encodings such as the multi-channel representation of quantum images~\cite{Sun2011, Sun2013}. This extension is crucial for applying QML to real-world computer vision tasks where chromatic features are essential for classification and segmentation.

More broadly, reducing the cost of state preparation is a necessary step toward making quantum computing practical for data-driven applications, since quantum algorithms can only deliver useful advantages when the cost of loading and representing classical data is itself scalable.


\bibliographystyle{IEEEtran}
\bibliography{references}
\end{document}